\documentstyle[twoside,fleqn,espcrc2,epsf]{article}
\begin{document}
% declarations for front matter
\title{The Flat Phase of Fixed-Connectivity Membranes\thanks{Combined
contribution of the talk presented by M.\ Bowick and the poster
presented by M.\ Falcioni.}}

\author{{\bf M.J.\ Bowick}$^{\rm a}$, S.M.\ Catterall$^{\rm a}$, {\bf
M.\ Falcioni}$^{\rm a}$, G.\ Thorleifsson\address{Department of
Physics, Syracuse University,\\ Syracuse, NY, 13244-1130, U.S.A.\ }
and K.\ Anagnostopoulos\address{The Niels Bohr Institute,\\
Blegdamsvej 17, DK-2100 Copenhagen \O, Denmark}}

\begin{abstract}
The statistical mechanics of flexible two-dimensional surfaces
(membranes) appears in a wide variety of physical settings.  In this
talk we discuss the simplest case of fixed-connectivity surfaces.  We
first review the current theoretical understanding of the remarkable
flat phase of such membranes. We then summarize the results of a
recent large scale Monte Carlo simulation of the simplest conceivable
discrete realization of this system \cite{BCFTA}.  We verify the
existence of long-range order, determine the associated critical
exponents of the flat phase and compare the results to the predictions
of various theoretical models.
\end{abstract}

\maketitle

\section{INTRODUCTION}
Physical membranes, or 2-dimensional surfaces embedded in {\bf
R}\(^3\), are believed to have a high-temperature crumpled phase and a
low temperature flat phase \cite{NP1}. The flat phase is characterized
by long range orientational order in the surface normals.  Since
long-range order is highly unusual in 2-dimensional systems, it is
worthwhile developing a thorough understanding of this phase.

There are several experimental realizations of crystalline surfaces.
Inorganic examples are thin sheets (\(\le 100\) \AA) of graphite oxide
(GO) in an aqueous suspension \cite{graphite1,graphite2} and the
rag-like structures found in MoS\(_2\) \cite{CPPDN}.

There are also remarkable biological examples of crystalline surfaces
such as the spectrin skeleton of red blood cell membranes. This is a
two-dimensional triangulated network of roughly 70,000
plaquettes. Actin oligomers form nodes and spectrin tetramers form
links \cite{Skel}. Crystalline surfaces can also be synthesized in the
laboratory by polymerising amphiphillic mono- or bi-layers. For recent
reviews see \cite{Peliti,fdavid,jer2}.

In this contribution we first review briefly the current analytical
understanding of rigid membranes and then summarize the results of a
recent large-scale Monte-Carlo simulation \cite{BCFTA}.

A Landau-Ginzburg-Wilson effective Hamiltonian for a \(D\)-dimensional
elastic manifold with bending rigidity, embedded in \(d\)-dimensional
space is
\begin{eqnarray}
\nonumber H_{\it eff} &=& \int {\rm d}^D \! \sigma \; \left[ \frac{\kappa}{2}
\left(\partial_\alpha \vec{\phi}_\alpha \right)^2 +
\frac{t}{2} \vec{\phi}_\alpha \cdot \vec{\phi}_\alpha \right.\\
& + &  \left. u \left( \vec{\phi}_\alpha
\cdot \vec{\phi}_\beta \right)^2 + v \left( \vec{\phi}_\alpha
\cdot \vec{\phi}_\alpha \right)^2\right],
\label{eq:effham}
\end{eqnarray}
where the order field \(\vec{\phi}_\alpha = \partial_\alpha \vec{r}\),
and \(\vec{r}(\sigma)\) is a vector in {\bf R}\(^d\).  We are clearly
dealing with a matrix \(\phi^4\) theory.  In the crumpled phase (\(t >
0\)) the position vector \(\vec{r}\) scales with system size \(L\)
like \(\vec{r} \sim L^\nu\), with \(\nu < 1\), and hence \(\phi_\alpha
\sim L^{\nu-1}\). The exponent \(\nu = 2/d_H\) is the size or Flory
exponent, where \(d_H\) is the Hausdorff dimension.  An expansion in
\(\phi_\alpha\) and its gradients is therefore justified.  In the flat
phase (\(t < 0\)), the Hamiltonian is stabilized by the anharmonic
terms.  In mean field theory one finds that the induced metric
\(g_{\alpha \beta}\) has non-zero expectation value, \(\langle
\partial_\alpha \vec{r} \cdot \partial_\beta \vec{r}\rangle \propto
\delta_{\alpha \beta}\).

\begin{figure}
\centerline{\epsfxsize=2.8in \epsfbox{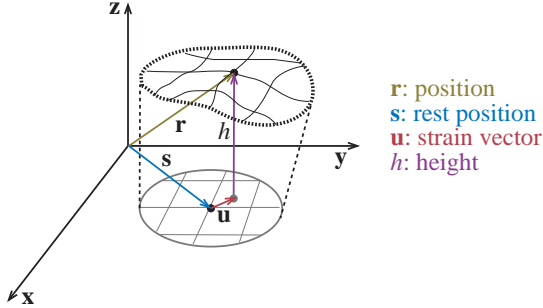}}
\caption{The Monge gauge.} 
\label{fig:monge}
\end{figure}

The Hamiltonian Eq.\ (\ref{eq:effham}) takes the following form in
the flat phase
\begin{equation}
 H_{\it eff} = \frac{1}{2} \int d^D \! \sigma \; \left[ \left(
\partial^2 \vec{r} \right)^2 + 2\mu \, u_{\alpha \beta}^2 + \lambda \,
u_{\gamma\gamma}^2\right],
\label{eq:flatham}
\end{equation}
where the strain tensor \(u_{\alpha\beta} = \frac{1}{2}
(\partial_\alpha \vec{r} \cdot \partial_\beta \vec{r} -
\delta_{\alpha\beta}) \). The upper critical dimension for this model
is 4 and it may be analyzed in a \(\epsilon = 4-D\) expansion.  The
most pressing physical question is the value of the lower critical
dimension \(D_{l}\).  In particular one would like to know if \(D_l
\leq 2\), in which case physical membranes would indeed have a stable
flat phase.  The Hamiltonian Eq.\ (\ref{eq:flatham}) and the strain
tensor \(u_{\alpha\beta}\) can be re-written in the Monge gauge (see
Figure \ref{fig:monge}) as
\begin{equation}
H_{\it eff} = \frac{1}{2} \int d^D \! \sigma \left[ \kappa\,
(\partial^2 h)^2 + 2\mu \, u_{\alpha \beta}^2 + \lambda \,
u_{\gamma\gamma}^2 \right]
\label{eq:monge}
\end{equation}
and
\begin{equation}
u_{\alpha\beta} =  \frac{1}{2}(\partial_\alpha u_\beta +
\partial_\beta u_\alpha + \partial_\alpha h \partial_\beta
h),
\end{equation}
where \(\mu\) and \(\lambda\) are the Lam\'e
coefficients.\footnote{Note that it is impossible for the surface to
trade stretching energy for bending energy so as to make \(u_{\alpha
\beta}\) everywhere zero, since the two phonon degrees of freedom are
insufficient to cancel the three independent components of the strain
tensor.}
\begin{table*}[hbt]
\setlength{\tabcolsep}{1.5pc}
\catcode`?=\active \def?{\kern\digitwidth}
\caption{Theoretical predictions and numerical results.}
\label{tbl:results}
\begin{tabular*}{\textwidth}{@{}l@{\extracolsep{\fill}}cccc}
\hline  
 & \(\zeta\)  & \(\eta_u\) & \(\nu = {2}/{d_H}\) & \(\eta\)\\ \hline 
AL & \({13}/{25}\) or 0.52 & \({2}/{25}\) or 0.08 & 1 &
 \({24}/{25}\) or 0.96 \\ 
Large-\(d\) & \({2}/{3}\) &\({2}/{3}\) & 1 & \({2}/{3}\) \\ 
SCSA & 0.590 & 0.358 & 1 & 0.821\\ 
MC & 0.64(2) & 0.50(1) & 0.95(5) & 0.62 \\ \hline \hline
\end{tabular*}
\end{table*}

By rescaling \(\sigma \to s \sigma\), one sees that Eq.\
(\ref{eq:monge}) in 4 dimensions is a function only of the
dimensionless parameters 
\begin{equation}
\tilde{\mu} = \frac{\mu}{\kappa^2} \;\; {\rm and} \;\;
\tilde{\lambda} = \frac{\lambda}{\kappa^2}.
\label{eq:scaling}
\end{equation}
Aronovitz and Lubensky (AL) determined the RG flow of \(\tilde{\mu}\)
and \(\tilde{\lambda}\) within the \(\epsilon\)-expansion, at fixed
co-dimension, and found a globally attractive IR-stable fixed point at
infinite bending rigidity \cite{AL}. This fixed point should control
the properties of the whole flat phase \cite{AGL,GDLP,GDLP2}. In
particular, we can introduce the anomalous scaling dimensions for the
running coupling constants, \(\kappa_R(q) \sim q^{-\eta}\) and
\(\mu_R(q) \sim \lambda_R(q) \sim q^{\eta_u}\). The generalization of
Eq.\ (\ref{eq:scaling}) to \(D\) dimensions relates the exponents
\(\eta\) and \(\eta_u\) with the scaling identity
\begin{equation}
\label{eq:scal}
\eta_u = 4 - D - 2 \eta.
\end{equation}
The roughness exponent \(\zeta\), which governs the scaling of the
height-height correlation function, is related to the exponent
\(\eta\) by the scaling relation \(\zeta = (4 - D - \eta)/2\).  One
may also study the Hamiltonian of Eq.\ (\ref{eq:monge}) in a
large-\(d\) expansion in the non-linear sigma model limit (infinite
Lam\'e coefficients) \cite{DG}. Finally one can solve a set of
self-consistent equations (SCSA) for the scaling exponents of the
renormalized coupling constants \(\kappa\), \(\mu\) and
\(\lambda\) \cite{LDR}. The predictions for \(\zeta\), \(\eta\),
\(\eta_u\) and \(\nu\) from the different approximations, and the
results of our Monte Carlo (MC) simulations are summarized in Table
\ref{tbl:results}.

\section{THE MODEL}
\begin{figure}
\centerline{\epsfxsize=2.8in \epsfbox{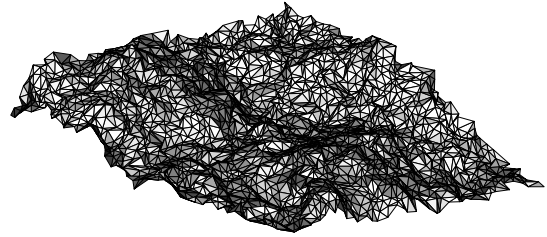}}
\caption{A snapshot of the surface in the flat phase for \(L=46\).} 
\label{fig:conf}
\end{figure}

In the simplest discretized version the crystalline surface is
modeled by a regular triangular lattice with {\em fixed}
connectivity, embedded in \(d\)-dimensional space.  Typically the
link-lengths of the lattice are allowed some limited fluctuations.
This is usually modeled by tethers between hard spheres or by
introducing some confining pair potential with short-range repulsion
between nodes (monomers), such as Lennard-Jones.  In some cases, the
bending energy is explicitly introduced, and is represented by a
ferromagnetic-like interaction between the normals to
nearest-neighbor ``plaquettes'' \cite{jer2}.

We are interested in a much simpler model of a crystalline surface,
inspired by the Polyakov action for strings \cite{Poly}. The tethering
potential between the particles is a simple Gaussian potential, with
vanishing equilibrium length.  Since the equilibrium length defines
the bare elastic constants of the surface one may wonder whether such
a model indeed exhibits the same phase diagram, and, in particular, if
it has a stable flat phase.

\(N\) particles are arranged in a regular triangular mesh. Each node
in the interior has 6 neighbors.  The action is composed of a
tethering potential and a bending energy term:
\begin{equation}
\label{eqn:action}
H = \frac{1}{2} \sum_{\langle ij \rangle} (\vec{r}_i - \vec{r}_j)^2 +
\frac{\kappa}{2} \sum_{\langle ab \rangle} (1 - \vec{n}_a \cdot
\vec{n}_b) 
\end{equation}
where \(\vec{r}\) is the position of node \(i\), and \(\vec{n}_a\) is
the unit normal to face \(a\).  The sums extend to nearest neighbors.
We point out that the normal-normal interaction translates to a
next-to-nearest neighbor interaction (like a \(\nabla^2\) term.)  We
do not include any minimum distance between the nodes. The model
describes a {\em phantom} surface, since there is no self-avoidance
term.  While self-avoidance changes the nature of the crumpling
transition, in the flat phase it is irrelevant (see Y.\ Kantor in
\cite{jer2}.)

We choose to simulate a surface with free boundaries, since it
simplifies the analysis of the correlation functions.  The trade off
is that we have to take careful account of edge fluctuations.  This is
described and illustrated in detail in \cite{BCFTA}.

\section{OBSERVABLES}
\begin{figure}
\centerline{\epsfxsize=2.8in \epsfbox{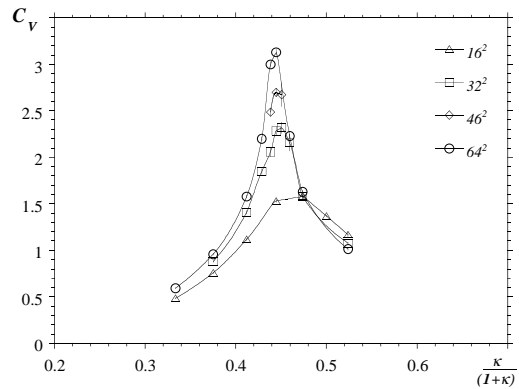}}
\caption{The specific heat \(C_V\) versus \(\kappa/(1 + \kappa)\).} 
\label{fig:spec}
\end{figure}
In order to estimate the exponents shown in Table \ref{tbl:results} we
measured several observables.  For finite sizes, the specific heat
\(C_V\) peaks in the vicinity of a second order phase transition, or
at coupling \(\kappa \simeq \kappa_c\).  The peak diverges with the
exponent \(\alpha \simeq 0.4\).  We are currently increasing the
statistics around the phase transition in order to determine
\(\alpha\) with greater accuracy.  The main uncertainty affecting the
estimate for \(\alpha\) comes from the estimate of \(\kappa_c\). We
also measure \(C_V\) in order to locate a region of the flat phase
suitable for studying its scaling behavior.  Note that, for a surface of
finite size, the bending rigidity controls the importance of finite
size effects. In order to obtain reliable finite size scaling one has
to tune the correlation length \(\xi \sim L\).

While \(C_V\) indicates the location of the transition, it tells
little about the nature of the phases on each side.  Thus we measured
the shape tensor and computed its eigenvalues.  The shape tensor is
the off-diagonal part of the inertia tensor \(I_{\alpha\beta}\), being
in some sense orthogonal to it.  It is defined as
\begin{eqnarray}
\label{eq:shape}
S_{\alpha\beta} &=& \left\langle \sum_\sigma r_\alpha(\sigma)
r_\beta(\sigma) \right\rangle_{\!c}\\ 
\nonumber & = & \delta_{\alpha\beta} \left\langle
\sum_\sigma r_\gamma(\sigma) r_\gamma(\sigma) \right\rangle_{\!c} -
I_{\alpha\beta}, 
\end{eqnarray}
where \(\alpha\), \(\beta\) refer to the components of \(\vec{r}\),
and the subscript \(c\) indicates connected expectation values.  The
radius of gyration \(R_g^2 = {\rm tr} S\) changes drastically
across the transition.  While for \(\kappa < \kappa_c\) (hot phase)
\(R_g\) has small values compared with the linear size \(L\) of the
surface, for \(\kappa > \kappa_c\) \(R_g\) is large.  More importantly
the finite size scaling of \(R_g \sim \L^\nu\) defines the size
exponent \(\nu = 2/d_H\). In the crumpled phase \(R_g \sim \log(L)\),
and \(\nu = 0\).  At the critical point \(\nu\) has a non-trivial
value.  Above the critical point, in the flat phase, \(\nu = 1\) or
\(d_H = 2\).  Our measurement of \(\nu = 0.95(5)\) at \(\kappa =
1.1\).
\begin{figure}
\centerline{\epsfxsize=2.8in \epsfbox{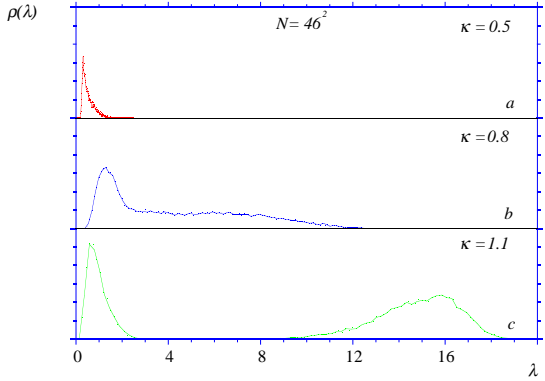}}
\caption{The distribution of eigenvalues of the shape tensor.} 
\label{fig:eigs}
\end{figure}

The eigenvectors of \(S\) define a body-fixed frame on the surface;
the eigenvalues of \(S\) are the average square dispersions in the
direction of the associated eigenvalue and tell more about the shape
of the surface.  Figure \ref{fig:eigs} shows the distribution of
eigenvalues \(\rho(\lambda)\) in the three regions of the phase
diagram.  Box a) shows \(\rho\) in the crumpled phase: all three
eigenvalues have identical distribution and the system is isotropic.
Box b) shows \(\rho\) in the vicinity of the transition: the system is
still isotropic but the eigenvalues have fluctuations on large scales.
Box c) shows \(\rho\) in the flat phase: the surface is no longer
isotropic --- there is a well defined thickness and lateral extension,
corresponding, respectively, to the left and right peak.

The roughness exponent \(\zeta\) is measured from the finite size
scaling of the average thickness of the surface. The minimum
eigenvalue of \(S\) provides just this information.  This particular
observable is very sensitive to boundary effects. In fact if an edge
of the surface is ``curled'', the eigenvalue will be considerably
larger even if the surface is locally quite smooth.  

\begin{figure}
\centerline{\epsfxsize=2.8in \epsfbox{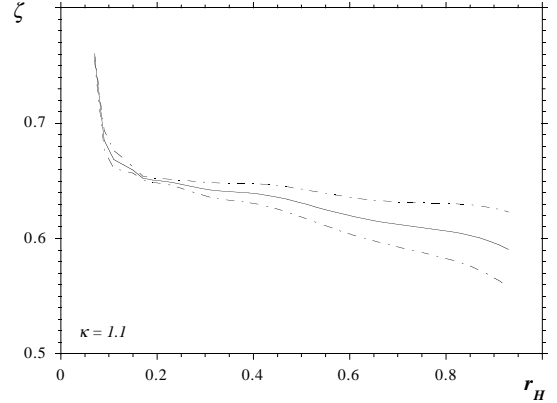}}
\caption{The roughness exponent \(\zeta\) versus the diameter of the
hexagonal sub-set of the surface.} 
\label{fig:zeta}
\end{figure}

In order to determine the influence of the boundary, we measured the
average square thickness of several concentric hexagonal sub-sets of
the surface of varying diameter \(D\).  We found that \(\zeta\)
plateaus in the range \(L/4 < D < 3L/4\), shows boundary effects for
\(D>3L/4\) and discretization effects for \(D<L/4\).  We quote the
value extracted from the plateau region (see Figure \ref{fig:zeta}).

The exponent \(\eta_u\) can be extracted from the {\em phonon
fluctuations}.  It is in fact straight-forward to relate this
observable and the exponent using the finite size scaling relation
\begin{equation}
\langle \vert \vec{u} \vert^2 \rangle \sim L^{\eta_u}.
\end{equation}
We have performed precise measurements of the phonon fluctuations,
described in detail in ref.\ \cite{BCFTA}.  The determination of this
exponent is important since it provides and independent
consistency check of the scaling relation (\ref{eq:scal}).  Our
measurement, shown in Table \ref{tbl:results}, is in good agreement
with the theoretical predictions and the scaling relation.
\begin{table*}[t]
\setlength{\tabcolsep}{1.5pc}
\newlength{\digitwidth} \settowidth{\digitwidth}{\rm 0}
\catcode`?=\active \def?{\kern\digitwidth}
\caption{The number of thermalized sweeps collected per data point in
the flat phase. The last column indicates the autocorrelation time
for the slowest mode in the system, the radius of gyration.}
\label{tbl:runs}
\begin{tabular*}{\textwidth}{@{}l@{\extracolsep{\fill}}rrr}
\hline \(L\) & \multicolumn{1}{r}{\(\kappa = 1.1\)} &
\multicolumn{1}{r}{\(\kappa = 2.0\)} & \multicolumn{1}{r}{\(\tau_R
\sim L^z\)} \\ \hline 32 & \(31 \times 10^6\) & \(26 \times 10^6\) &
\(3 \times 10^4\) \\ 46 & \(51 \times 10^6\) & \(42 \times 10^6\) &
\(7 \times 10^4\) \\ 64 & \(47 \times 10^6\) & \(44 \times 10^6\) &
\(1.2 \times 10^5\)\\ 128 & \(74 \times 10^6\) & --- & \(5 \times
10^5\)\\ \hline \hline
\end{tabular*}
\end{table*}

Our final measurement of the properties of the flat phase of this
model is the normal-normal correlation function.  We expect the
correlation function to fall off to a non-zero asymptote like
\begin{equation}
\label{eqn:nnfalloff}
\langle \vec{n}_\sigma \cdot \vec{n}_o\rangle \; \sim \; C +
\frac{c}{r^\eta},
\end{equation}
where \(r\) is the geodesic distance between the center \(o\) and the
point \(\sigma\).  Since the boundaries are free, the correlation
function is not translationally invariant: we therefore fix the origin
at the center of the surface and discard data too close to the
boundary.  
\begin{figure}
\centerline{\epsfxsize=2.8in \epsfbox{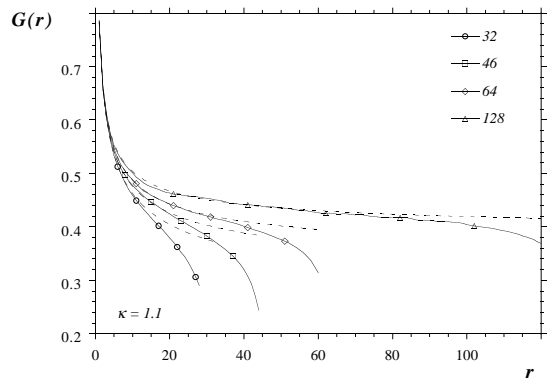}}
\caption{The normal-normal correlation function for different lattice
sizes in the flat phase.  The dashed lines are our best fit to Eq.\
\ref{eqn:nnfalloff}.} 
\label{fig:nn}
\end{figure}

In Figure \ref{fig:nn} we show the behavior of the correlation
function for different values of the lattice size.  From the data in
the figure it is clear that the normal-normal correlation function has
a non-zero asymptote.  The value of the asymptote grows with system
size, and we believe it has a non zero infinite volume limit.  The fit
of the data to Eq.\ \ref{eqn:nnfalloff} gives yet another independent
estimate of the exponent \(\eta\).  Our best fit to the \(L=128\) data
gives \(\eta \simeq 0.62\), in good agreement with our previous
estimates.

In conclusion, we have shown that the simple gaussian model of Eq.\
\ref{eqn:action} faithfully reproduces the expected critical behavior
of the AL fixed point.  The relative simplicity of this model enables
us to simulate surfaces of realistic size. The spectrin network of a
blood cell has about 70,000 plaquettes; our largest surface has 32,768
plaquettes.  

\section{NUMERICAL METHODS}
We will now turn our attention to the technical
aspects of the simulation methods and the computers used.  We use a
Monte Carlo algorithm with a local Metropolis update.  We choose a new
position for a given node in a box of size \(\epsilon^3\) centered on
the old position, and we accept it according to the standard
Metropolis test.  A single Monte Carlo sweep consists of an update of
all the nodes of the surface.  The box size \(\epsilon\) is adjusted
so as to keep the acceptance ratio around 50\%.

\begin{table*}[t]
\setlength{\tabcolsep}{1.5pc}
\catcode`?=\active \def?{\kern\digitwidth}
\caption{Wall-clock timing results of the benchmark runs for different
serial and parallel codes.  The number are in seconds.}
\label{tbl:bench}
\begin{tabular*}{\textwidth}{@{}l@{\extracolsep{\fill}}rrrrr}
\hline 
 & \multicolumn{1}{c}{Serial code}& \multicolumn{2}{c}{MPI code} &
\multicolumn{2}{c}{HPF code} \\ \cline{2-2} \cline{3-4} \cline{5-6} 
& \multicolumn{1}{c}{A} 
& \multicolumn{1}{c}{A}
& \multicolumn{1}{c}{B} 
& \multicolumn{1}{c}{A}
& \multicolumn{1}{c}{B}  \\ \hline
Serial & 334 & --- & --- & 4800 & --- \\
2      & --- & 705 & 1410 & 2189 & 4378 \\
4      & --- & 400 & 1600 & 2375 & 9500 \\
8      & --- & 212 & 1696 & --- & ---\\ 
 \hline \hline
\end{tabular*}
\end{table*}

Crystalline surfaces are characterized by extremely long
autocorrelation times.  The modes which suffer most are the ones
related to the global {\em shape} of the surface in the embedding
space.  In table \ref{tbl:runs} we show the amount of data collected
at various sizes/bending rigidities, and the corresponding
autocorrelation time for the radius of gyration.  We estimate the
critical slowing down exponent \(z\) to be \(\simeq 2\), as expected
for a local algorithm.

We are currently investigating several ways of improving the 
Monte Carlo simulations methods.  In particular, we are
studying different algorithms, like over-relaxation and uni/multi-grid
Monte Carlo.  Previous numerical studies have successfully taken the
advantage of Fourier acceleration \cite{BET,ET2,Wheater}, and it is
argued that multi-grid methods should perform similarly well
\cite[p.\ 41]{Sokal:1989}.  Preliminary tests of these algorithms
give very encouraging results: over-relaxation reduces significantly
the autocorrelation time, while we expect the multi-grid to improve
the dynamical exponent \(z\).

We used several workstations for most of the runs, but we simulated
the largest lattice size (\(L=128\)) using a MASPAR MP1, a massively
parallel processor, with exactly 16,384 nodes.  Due to its hardware,
the MP1 is best suited to run a Monte Carlo simulation with local
update.  Massively parallel processors are no longer common:
workstation clusters with high speed switches are becoming more and
more popular.  This trend is reflected in the parallelization
strategies that we are currently investigating.

While High Performance Fortran (HPF) is clearly the better choice for
massively parallel computers, such as the MP1, it is still unclear
whether its efficiency (or lack thereof) justifies its use in
clustered environments.  The alternative is to use the Message Passing
Interface (MPI). This is a library of functions and a set of
programming models which allows one to distribute a simulation over a
cluster of workstations.

It is much simpler to write programs in HPF than MPI, since the HPF
compiler automatically distributes the data on the various processors
and it introduces the appropriate parallel instructions.  The
trade-off is that the programmer has little control over how the
parallelization is actually done, and often the compiler does not
exploit all of the potential parallelism of the algorithm.  MPI allows
much more flexibility and tailoring, but it is much harder to use
since the parallelization must be coded by hand.

We have performed a series of benchmarks, using a simplified version
of the code, in order to establish the efficiency of the different
parallelization techniques.  In Table \ref{tbl:bench} we compare the
results of the benchmark runs.  The first column shows the number of
processors used in the run. This is an important factor: more
processors mean more computing power, but also more message passing
between the processors.  While massively parallel computers had
relatively low latency times for communication, clustered workstations
often rely on a common switchboard with high latency times.  The first
column lists the elapsed time for a single processor run of the scalar
code --- this is the reference.  The third and fourth column have the
timings for the MPI code, written in {\tt c}.  Column A shows the
elapsed time, while column B shows the integrated time (wall-clock time
\(\times\) number of processors).  The fifth and sixth column show the
timing of the HPF code.  We note that, while for the MPI code the
integrated time (column 3) grows slowly with the number of processors,
for the HPF code (column 6) the time {\em doubles}, indicating very
inefficient message passing.  Based on our limited experience with our
particular code, we believe that HPF is still far from being a viable
choice for parallel programming on clustered environments.

Finally, we believe that farming (or running independent serial
simulations on many processors) is still the most efficient solution
for simulations small enough to fit on a single processor, since the
integrated time for the MPI code is still higher than the single
processor time.

\section{ACKNOWLEDGMENTS}
We would like to thank David Nelson, Mehran Kardar, Emmanuel Guitter,
Alan Middleton, Paul Coddington, Enzo Marinari and Gerard Jungman for
helpful discussions.  NPAC (North-East Parallel
Architecture Center) has provided the computational facilities.
The research of MB and MF was supported by the Department of Energy
U.S.A.\ under contract No.\ DE-FG02-85ER40237.  SC and GT were
supported by research funds from Syracuse University.  Part of the
work of KA was done at the Institute for Fundamental Theory at
Gainesville and was supported by DOE grant No.\ DE-FG05-86ER-40272. We
also acknowledge the use of the software package Geomview for membrane
visualisation and code development \cite{Geomview}.

\bibliography{paper}

\begin{thebibliography}{10}

\bibitem{BCFTA}
M.~J. Bowick {\it et~al.}, The flat phase of crystalline membranes, to appear
  in J. Phys. I (France).

\bibitem{NP1}
D. Nelson and L. Peliti, J. Phys. France {\bf 48},  1085  (1987).

\bibitem{graphite1}
X. Wen {\it et~al.}, Nature {\bf 355},  426  (1992).

\bibitem{graphite2}
T. Hwa, E. Kokufuta, and T. Tanaka, Phys. Rev. A {\bf 44},  R2235  (1991).

\bibitem{CPPDN}
R.~R. Chianelli, E.~B. Prestridge, T. Pecoraro, and J.~P. DeNeufville, Science
  {\bf 203},  1105  (1979).

\bibitem{Skel}
C.~F. Schmidt {\it et~al.}, Science {\bf 259},  952  (1993).

\bibitem{Peliti}
L. Peliti,  in {\em Fluctuating Geometries in Statistical Mechanics and Field
  Theory}, edited by P. Ginsparg, F. David, and J. Zinn-Justin ({\tt
  cond-mat/9501076}, Les Houches, France, 1994), lectures given at the Les
  Houches Summer School.

\bibitem{fdavid}
F. David,  in {\em Two Dimensional Quantum Gravity and Random Surfaces}, Vol.~8
  of {\em Jerusalem Winter School for Theoretical Physics}, edited by D. Gross,
  T. Piran, and S. Weinberg (World Scientific, Singapore, 1992).

\bibitem{jer2}
{\em Statistical Mechanics of Membranes and Surfaces}, Vol.~5 of {\em Jerusalem
  Winter School for Theoretical Physics}, edited by D. Nelson, T. Piran, and S.
  Weinberg (World Scientific, Singapore, 1989).

\bibitem{AL}
J. Aronovitz and T. Lubensky, Phys. Rev. Lett. {\bf 60},  2634  (1988).

\bibitem{AGL}
J. Aronovitz, L. Golubovi\'c, and T. Lubensky, J. Phys. France {\bf 50},  609
  (1989).

\bibitem{GDLP}
E. Guitter, F. David, S. Leibler, and L. Peliti, Phys. Rev. Lett. {\bf 61},
  2949  (1988).

\bibitem{GDLP2}
E. Guitter, F. David, S. Leibler, and L. Peliti, J. Phys. France {\bf 50},
  1787  (1989).

\bibitem{DG}
F. David and E. Guitter, Europhys. Lett. {\bf 5 (8)},  709  (1988).

\bibitem{LDR}
P. {Le~Doussal} and L. Radzihovsky, Phys. Rev. Lett {\bf 69},  1209  (1992).

\bibitem{Poly}
A. Polyakov, Nucl. Phys. B {\bf 268},  406  (1986).

\bibitem{BET}
M. Baig, D. Espriu, and A. Travesset, Nucl. Phys. B {\bf 426},  575  (1994).

\bibitem{ET2}
D. Espriu and A. Travesset, Nucl. Phys. B (Proc. Suppl.) {\bf 47},  637
  (1996).

\bibitem{Wheater}
J. Wheater, Nucl. Phys. B {\bf 458},  671  (1996).

\bibitem{Sokal:1989}
A.~D. Sokal, {Monte Carlo} mehtods in statistical mechanics: foundations and
  new algorithms, given at the Troisieme Cycle de la Physique en Suisse
  Romande, Lausanne, Switzerland, Jun 15-29, 1989.

\bibitem{Geomview}
M. Phillips, S. Levy, and T. Munzner, Notices of the American Mathematical
  Society  985  (1993), computers and mathematics column.

\end{thebibliography}
\end{document}